\documentstyle[prb,aps]{revtex}
\begin{document}
\draft

\title{Direct simulation of ion beam induced stressing
and amorphization of silicon}

\author{Keith M. Beardmore}
\address{Motorola Computational Materials Group,
Los Alamos National Laboratory, Los Alamos, NM 87545}

\author{Niels Gr{\o}nbech-Jensen}
\address{Theoretical Division,
Los Alamos National Laboratory, Los Alamos, NM 87545}

\date{\today; first submitted Jan 27, 1999}
\maketitle

\begin{abstract}
Using molecular dynamics (MD) simulation, we investigate the
mechanical response of silicon to high dose
ion-irradiation.
We employ a realistic and efficient model to directly
simulate ion  beam induced amorphization.
Structural properties of the amorphized sample are compared with
experimental data and results of other simulation studies.
We find the behavior of the irradiated material is related to the rate at
which it can relax.
Depending upon the ability to deform, we observe either the generation
of a high compressive
stress and subsequent expansion of the material,
or generation of tensile stress and densification.
We note that statistical material properties, such as radial distribution
functions are not sufficient to differentiate between
different densities of amorphous samples.
For any reasonable deformation rate, we observe an
expansion of the target upon amorphization in agreement with
experimental observations.
This is in contrast to simulations of quenching which
usually result in denser structures relative to crystalline Si.
We conclude that although there is substantial agreement between experimental
measurements and most simulation results, the amorphous structures being
investigated may have fundamental differences;
the difference in density can be attributed
to local defects within the amorphous network.
Finally we show that annealing simulations of our amorphized samples can lead
to a reduction of high energy local defects without a large scale
rearrangement of the amorphous network. This supports the proposal
that defects in amorphous silicon are analogous to those in crystalline silicon.
\end{abstract}


\section{Introduction}
Mechanical response of materials to ion irradiation
has implications for many materials applications.
Ion processing of silicon is of widespread fundamental and technological
importance due to its central role in the semiconductor industry.
The high dose, large atomic number wafer
implants used today lead to a significant
amount of amorphization, which must be removed by
subsequent annealing.
Ion implant induced stresses can lead to
such problems as substrate bending, delamination and cracking,
and anomalous diffusion of dopants\cite{vol91}.
The creation and modification of stress in silicon by ion irradiation
has been examined experimentally
by several research groups\cite{vol91,fab89}.
Experimental measurements have shown that Si {\it expands} by approximately
1.8\% upon ion beam induced amorphization\cite{cus94}.
Little atomic scale modeling of the amorphization process has been conducted,
although many attempts have been made to simulate the structure of
amorphous Si ({\it a}-Si).
These simulations usually involve a rapid quench of liquid Si and suffer
from drawbacks such as limited system size and short simulation times. The
vast majority of these simulations result in a structure with
similar properties to experimental {\it a}-Si,
except for the fact that the density is
typically around 4\% {\it greater} than crystalline Si ({\it c}-Si)\cite{cus94}.
Here, we study the formation and structure of {\it a}-Si by direct
MD simulation of the ion beam amorphization process,
rather than by a simulation of quenching.

\section{Molecular Dynamics Simulation Scheme}
We employ a minimal atomistic model of the radiation damage process
and investigate stressing and deformation of the substrate.
A full MD simulation incorporating a large enough section of the
target material to completely contain the paths of many implanted ions,
and to run for a realistic
time is neither feasible, nor particularly desirable.
What we wish to study is the
response of a section of the material to the damage induced by the irradiation.
In order to achieve this, we simulate a model system, that
incorporates all the necessary features of the complete system.

At the ion energies used, the material response is a bulk,
rather than a surface effect, so  it is not necessary to explicitly
include the surface in simulations. Hence, periodic boundary conditions
are applied in all directions to minimize finite-size effects.
  Energy dissipation and equilibration of pressure are controlled by
coupling the system to an external bath\cite{ber84}. This involves
two parameters, $\tau_T$ and $\tau_P$, that are time constants controlling
the rate at which temperature and pressure tend towards set values.
   This non-physical coupling gives us control over quenching and expansion
of the system, but does limit the inferences we can draw about dynamic
processes. We have analyzed the dependence of simulation results on
the strength of these couplings in order to ensure that we do not introduce
any systematic errors into our simulations.
  We assume that the ion dose is sufficiently low that the presence of
ions in the material has a negligible effect and may be ignored.
The effect of the irradiation is modeled by choosing
primary knock-on atoms (PKAs) within the simulated region and assigning
velocities
to them from a distribution that can be dependent upon ion type, ion energy,
and the depth in the substrate.
PKAs are produced by collisions with the implanted ion, but the majority
are due to collisions of knock-on atoms in sub-cascades. Hence our PKA
velocity distribution must take account of all knock-on atoms that are
created by ion implantation.
In order to account for the free surface, 
the material is allowed to expand or contract in the direction normal
to the surface; expansion is not allowed in the in-plane directions.
   The passage of high energy ions through the simulated region are rare
events, between which the material will relax as energy is dissipated, then
enter an essentially constant state for a long period of time. As this constant
state contributes nothing to the response of the material, we omit it and
start another PKA at this time.
By taking
this approach, we can reduce the system size to the extent that a
representative set of simulations can be conducted on a few workstations
in a matter of days.

\section{Simulation Details}
MD based ion implant simulations\cite{reed} were used to
to determine realistic PKA velocity distributions for various
ion types and energies, at differing substrate depths; the resulting
distributions were similar.
The magnitude of PKA velocities could be approximately described by a
$\chi^2$ distribution. For an incident ion beam perpendicular to the surface,
the in-plane distribution of PKA velocity vectors was uniform.
The 
angle between the PKA velocity vector and the ion beam could be approximated
by a Gaussian distribution
with mean and standard deviation of $85^\circ$ and $10^\circ$ respectively.

PKA energies were chosen such that the size of displacement cascades was
less than the system dimensions,
to ensure that parts of the same damage cascade could not interact
through the periodic boundaries.
 The selection of coupling constants for pressure and temperature
control is critical; too long coupling times lead to excessive cpu
requirements, whilst too short coupling times lead to unphysical system
behavior, e.g., rapid quenching of hot disordered material will produce
an excessive number of defects.
Therefore, preliminary simulations were conducted
with various combinations of PKA energy, system size
and temperature and pressure coupling constants to investigate
and minimize the sensitivity of the simulation on these parameters.

The final simulations were conducted using PKA energies
with a mean of 1 keV.
All silicon atoms were coupled to a 300~K heatbath with a coupling constant,
$\tau_T$,
of 1.0 ps; this was determined to be large enough that increasing it
had no effect on simulation results.
A new PKA was initialized as soon as the system temperature was quenched
to below 320~K.
Stress in the direction normal to the surface was relaxed to zero
with a time constant, $\tau_P$, of
0.1, 1.0, or 10.0 ps.
No single value of this parameter could be used
as it effectively represents the ability of the material to locally
deform, both elastically and through plastic flow.
As such it is affected by the size of the ion path, ion mass and energy,
the ion dose rate and the total ion dose.
We therefore ran three sets of simulations spanning two orders of magnitude
range in this parameter. This range is expected to contain most experimentally
accessible conditions.
  In order to relate stress to change in atomic
positions, a Young's modulus of 113.0 GPa\cite{jan89} was used for both
crystalline and damaged Si;
the value of this parameter is expected to vary during
amorphization, but its exact value is not critical as variations will
only affect the value of the the non-critical time constant,
$\tau_P$\cite{ber84}.
The initial system was crystalline Si, with dimensions of approximately
50.0 \AA\ $\times$ 50.0 \AA\ $\times$ 50.0 \AA, containing 6240 atoms,
at zero pressure.
In order to be able to simulate a sufficiently large sample,
we use a many-body empirical potential for silicon\cite{ter88}.
To our knowledge, the derivation of the pressure virial for this
potential has not been previously published. A method of obtaining
the virial,
which may be applied to any many-body potential, is given in
Appendix\ \ref{app1}.

\section{Amorphization of silicon during ion irradiation}
Three sets of simulations are presented, the only difference between
them being the magnitude of the pressure control coupling constant.
In each case 9 separate simulations were conducted for each parameter set
and the final results averaged.
Data was recorded at the end of each quench, i.e.,
immediately prior to initializing a PKA,
to generate radial distribution functions (RDFs), and
distributions of bond angles, bond lengths, and atomic coordination.
   The simulation parameters chosen resulted in a mean time between
PKA initialization of approximately 6.25 ps.

\begin{figure}
\vbox{
\includegraphics{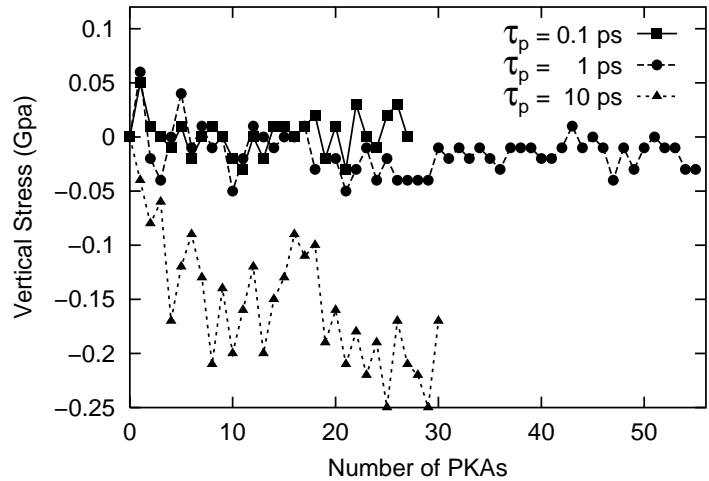}
\vspace{2.4 in}}

\caption{Variation in normal   stress during amorphization.}
\label{stress}

\end{figure}

\begin{figure}
\vbox{
\includegraphics{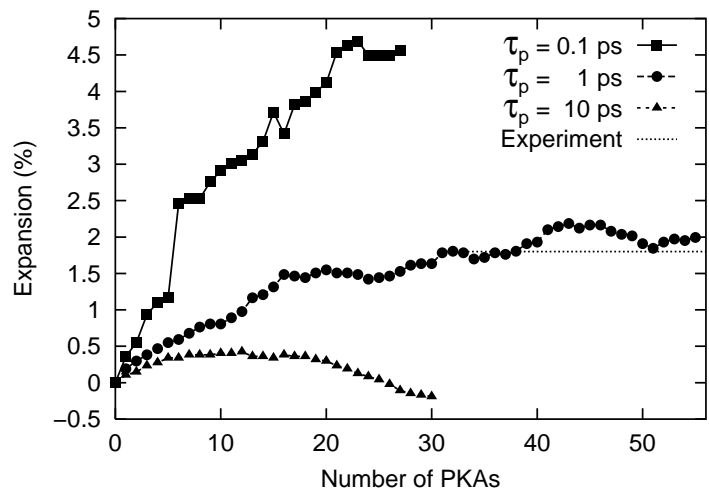}
\vspace{2.4 in}}

\caption{Change in height of simulation cell during amorphization.}
\label{height}

\end{figure}

During irradiation, damage accumulates within the sample, resulting in
the development of stress. As the simulation progresses, the material
expands, or contracts to relieve the normal   stress, as shown in
Figs.\ \ref{stress} to\ \ref{height}.
Anisotropic deformation occurs by flow of material
to relieve the in-plane stress.
The normal   stress is relaxed to zero
in the case of the two shorter $\tau_P$ values and remains negative for
the longer $\tau_P$.
The stress would be expected to reach an equilibrium state for any
value of $\tau_P$, given a sufficiently long simulation time,
but cpu constraints prevented us from achieving this for the largest value
of $\tau_P$.
The in-plane stress is intially positive (compression) due to damage
accumulation. Once a sufficient density of damage is present the sample
is able to relax via flow of material and the stress becomes
negative (tension).
The in-plane negative stress is due to the outflow of material during the
`thermal spike' phase of radiation damage. As the sample cools during
expansion a tensile state is reached, but material is not able to flow
back to relieve the stress.
 A similar response to ion irradiation has been previously
observed in wafer curvature experiments\cite{vol91}. At long times,
the damage appears to saturate and the material maintains a dynamic
equilibrium, with a constant in-plane stress of around $-0.3$~GPa.
The percentage expansion of the system is
dependent upon the value of $\tau_P$; smaller values (faster relaxation)
result in expansion of the material, whilst the largest value gives
an initial expansion followed by contraction. Due to cpu time constraints
only the intermediate case was followed until the system reached an
equilibrium condition. In this case the material reaches a density
around 1.9\% less than {\it c}-Si, approximately the same as
the experimental value of 1.8\%\cite{cus94}. 

\begin{figure}
\vbox{
\includegraphics{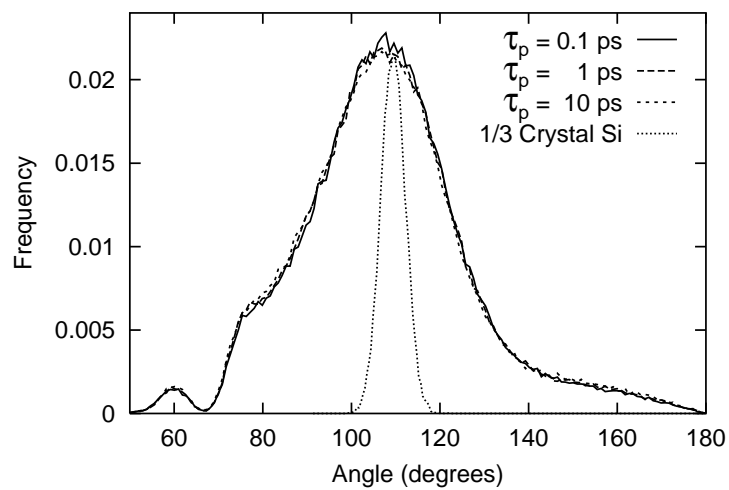}
\vspace{2.4 in}}

\caption{Final bond angle distributions in amorphized samples.}
\label{angle}

\end{figure}

\begin{figure}
\vbox{
\includegraphics{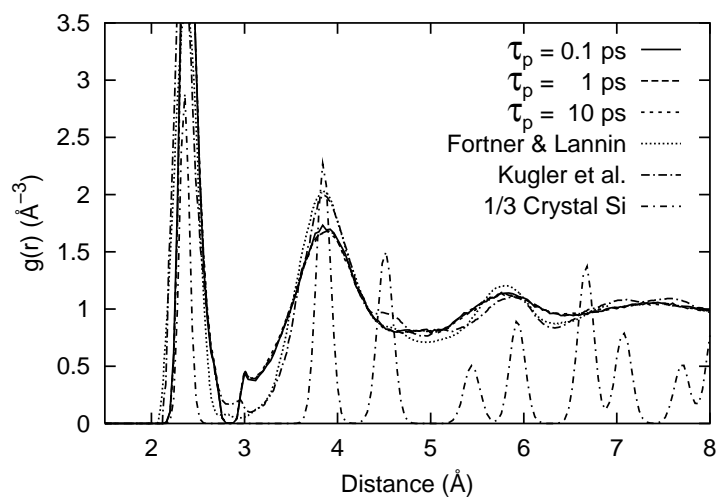}
\vspace{2.4 in}}

\caption{Final radial distribution functions in amorphized samples.}
\label{rdf}

\end{figure}

The distributions shown in Figs.\ \ref{angle} to\ \ref{rdf} were calculated
for the final state of each amorphization simulation, i.e., they correspond
to the final data points in Figs.\ \ref{stress} to\ \ref{height}.
The distribution of bond angles is shown in Fig.\ \ref{angle}. To calculate
the distribution, angles were weighted by the product of the Tersoff
cut-off functions of the two bonds involved. The small
peak at $60^\circ$ corresponds to three-fold rings and five-fold
coordinated atoms, the shoulder at around
$80^\circ$ is due to four fold rings, and the main peak is due to five and
six fold rings. The distribution due to {\it c}-Si is also shown to
illustrate the effect of damage and amorphization; the frequency is scaled
by $1/3$ in order to allow the sharp peak to appear
on the same plot as the {\it a}-Si
results. The mean bond angle in the amorphized samples is around $2^\circ$
less than the tetrahedral angle, with a standard deviation of approximately
$19^\circ$.

\begin{figure}
\vbox{
\includegraphics{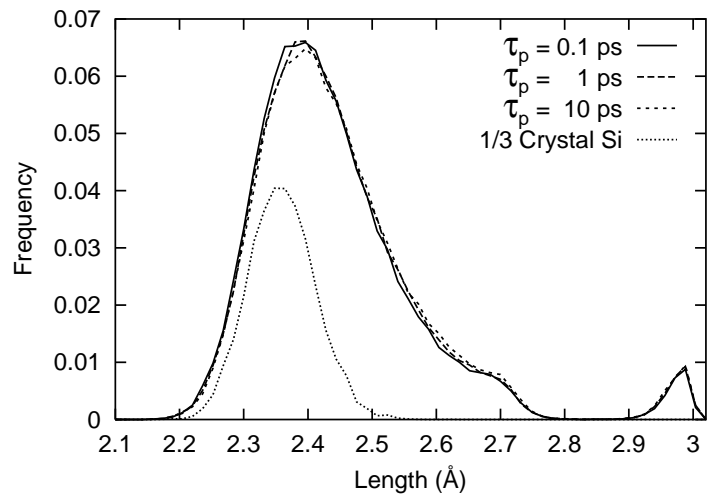}
\vspace{2.4 in}}

\caption{Final bond length distributions in amorphized samples.}
\label{length}

\end{figure}

\begin{figure}
\vbox{
\includegraphics{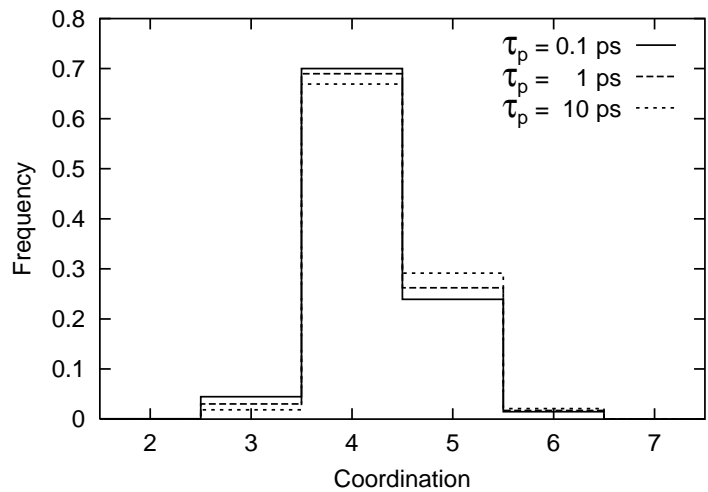}
\vspace{2.4 in}}

\caption{Final distributions of atomic coordination in amorphized samples.}
\label{coord}

\end{figure}

The distribution of bond lengths is shown in Fig.\ \ref{length}. When
interpreting this it is useful to recall that the Tersoff switching function
acts between 2.7 and 3.0 \AA. The small peak between 2.9 and 3.0 \AA\ is
therefore due to non-bonding (repulsive) interactions.
Bond lengths become about 0.09 \AA\ greater than those in {\it c}-Si.
The angle and length distributions due to the different relaxation
rates are so similar that they cannot be easily distinguished.
  In order to count the coordination of atoms, we have to specify a criterion
to decide if any pair of atoms are bonded. Based on the bond-length distribution
we count all atoms with interaction distances within the mid point of the
switching function (2.85~\AA) to be bonded. The resulting coordination
distribution is shown in Fig.\ \ref{coord}. Very few ($\sim$3\%)
undercoordinated atoms
are produced during amorphization, but approximately 26\% become
five-fold coordinated, and occasional ($\sim$2\%) six-fold coordination
occurs.
   Even though the average coordination of atoms increases to
approximately 4.27 during
amorphization, expansion is possible as bond-lengths also increase.

Table\ \ref{props} contains structural information on the samples produced
by these simulations.
The mean and standard deviation of the structural parameters are given,
and the percentage expansion of
the irradiated samples and potential energy
per atom are given relative to {\it c}-Si at 300 K.
The Tersoff potential gives
a cohesive energy for Si of 4.63 eV/atom at zero K and 4.59 eV/atom at 300 K
(this is in agreement with equi-partition of kinetic and potential energy,
as 300 K corresponds to a kinetic energy of 0.04 eV/atom).
   In the final damaged structures,
the potential energy per atom is approximately
0.41 eV higher than that of {\it c}-Si at 300 K. This stored energy is
approximately 8.9\% of {\it c}-Si cohesive energy and is primarily due
to bond angle distortions of around $19^\circ$ as shown in Fig.\ \ref{angle}
and Table\ \ref{props}.

The resulting radial distribution functions, shown in Fig.\ \ref{rdf}
are indistinguishable between
the various simulation sets, but differ from the experimental data
obtained by neutron diffraction experiments\cite{for89,kug89}.
The broader peaks in the simulated RDFs indicate that the structures contain
a higher degree of disorder than the experimental samples. There is also
a feature at around 3 \AA,
where the artifact in the shape of the first minimum
reflects the abrupt cutoff in the empirical potential.
The first two peaks of the {\it a}-Si RDF correspond to those for {\it c}-Si
showing the existence of short-range order and a tetrahedral like environment
for each atom. There is little correlation between the positions of the other
peaks, except for the third peak in the experimental data of Kugler et al.;
this may be an
indication of the presence of micro-crystallites in their sample.

\section{Annealing}
The damage simulations were carried out at room temperature with extremely
limited times for structural relaxation. Therefore, the amorphized samples
are expected to contain a large number of high energy defects; this is
demonstrated by the difference between the experimental and simulated RDFs.
In order to directly compare experimental and simulated samples, we have
therefore carried out annealing simulations. The final structures from each
set of damage simulations were taken, and subjected to four separate
anneals. Anneals were done at room temperature, between room
temperature and the {\it a}-Si to {\it c}-Si transition temperature,
between the transition temperature and the {\it c}-Si melting point,
and at the melting point. The Tersoff potential over-predicts the {\it c}-Si
melting temperature, giving a value of approximately 3000 K\cite{ter88}.
Therefore we applied the following scaling between actual and
simulated temperatures:
\begin{eqnarray}
T_{Tersoff} = \cases{T_{Experiment}      &$T_{Experiment} \le 300\ {\rm K}$\cr\cr
                {1.947\ T_{Experiment} - 283.994\ {\rm K} } &$T_{Experiment} > 300\ {\rm K}$.\cr}  \nonumber
\end{eqnarray}

This resulted in annealing temperatures of 300 K, 1121 K, 2194 K, and 3000 K,
which approximately correspond to experimental temperatures of
300 K, 722 K, 1273 K, and 1687 K
respectively.
Anneals were continued until no further change in any of the
measured structural parameters was observed.
All simulation parameters were kept the same as those
used for the damage simulations apart from the anneal temperature.
After annealing, the samples were quenched to 300 K over a time of 40 ps
before the final structural properties were measured.

\begin{figure}
\vbox{
\includegraphics{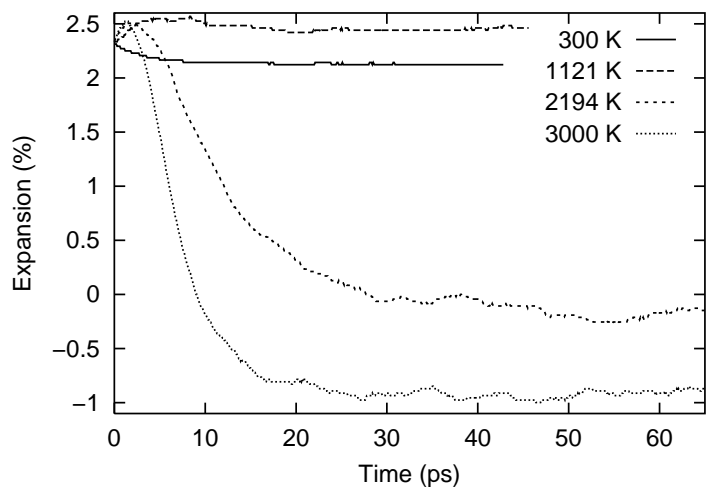}
\vspace{2.4 in}}

\caption{Change in height of simulation cell during annealing.}
\label{aheight}

\end{figure}

The structural changes during annealing are shown in the following plots
for the $\tau_P$=1 ps damage simulation.
The high temperature anneals resulted in large density changes as shown in
Fig.\ \ref{aheight}.
After cooling to room temperature to remove the effect of thermal expansion,
the final densities of the annealed samples corresponded to
2.1\% and 1.9\% expansion, and 0.4\% and 1.0\% contraction relative to
{\it c}-Si respectively for the 300 K, 1121 K, 2194 K, and 3000 K anneals.
   The room temperature anneal gives very little structural modification in
the time simulated; all distributions are indistinguishable from the
unannealed sample.

\begin{figure}
\vbox{
\includegraphics{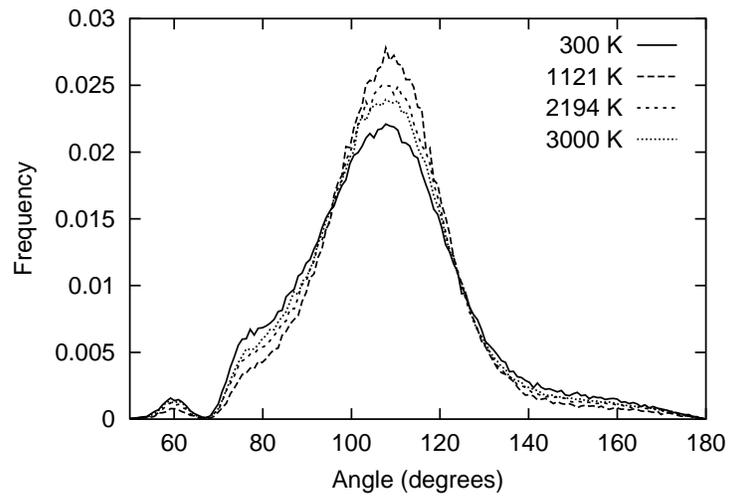}
\vspace{2.4 in}}

\caption{Final bond angle distributions in annealed samples.}
\label{aangle}

\end{figure}

\begin{figure}
\vbox{
\includegraphics{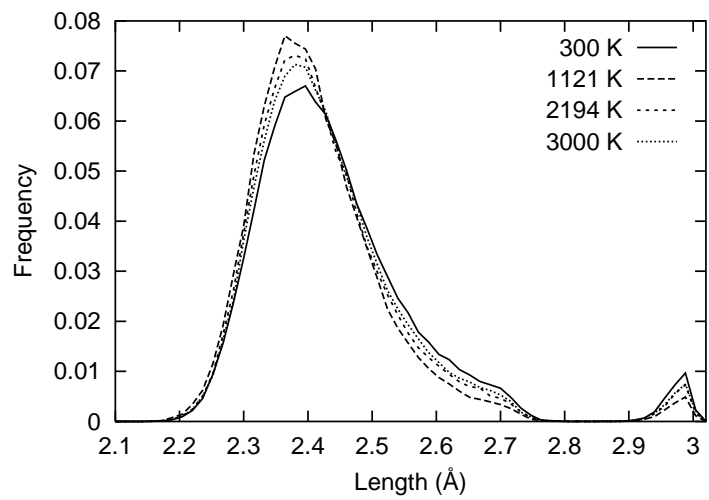}
\vspace{2.4 in}}

\caption{Final bond length distributions in annealed samples.}
\label{alength}

\end{figure}

\begin{figure}
\vbox{
\includegraphics{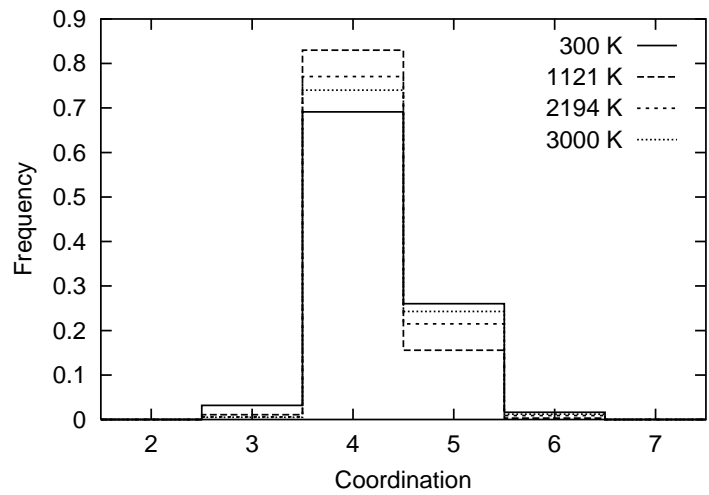}
\vspace{2.4 in}}

\caption{Final distributions of atomic coordination in annealed samples.}
\label{acoord}

\end{figure}

\begin{figure}
\vbox{
\includegraphics{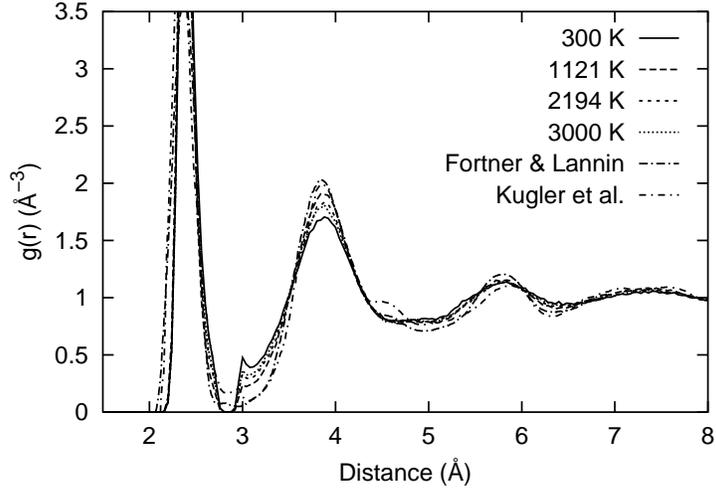}
\vspace{2.4 in}}

\caption{Final radial distribution functions in annealed samples.}
\label{ardf}

\end{figure}

The 1121 K anneal results in modifications to bond angle, bond length,
and coordination distributions and to the RDF, whilst the density is 
almost unchanged. This indicates that local defects are being removed
without any global structural reorganization.
This is in agreement with experimental studies which indicate that 
defects in {\it a}-Si are similar to defects in {\it c}-Si, i.e., they
behave as interstitials and vacancies within the amorphous
network\cite{roo91,lia94}.
Annealing therefore occurs through diffusion and annihilation of defects,
rather than by a
reorganization of the network.
   Fig.\ \ref{aangle} shows the final distribution of bond angles;
the mean increases by approximately
$1^\circ$, to around $1^\circ$ less than the {\it c}-Si angle, while the
standard deviation of bond angles is reduced to $16^\circ$.
   Bond lengths are reduced by an average of 0.03 \AA, to 0.06 \AA\ longer
than those in {\it c}-Si, as shown in Fig.\ \ref{alength}.
The anneal results in a reduction in 3-fold and 4-fold rings as evidenced
by the reduction in the frequency of bond angles of $60^\circ$
and $80^\circ$, and the reduction in 5-fold coordinated atoms.
  After annealing and quenching, approximately 1\% of atoms are 3-fold
coordinated, 83\% of atoms are 4-fold coordinated, and 16\% are 5-fold
coordinated, as shown in Fig.\ \ref{acoord}.
  The calculated radial distribution function, shown in Fig.\ \ref{ardf},
is in very good agreement with the experimental data,
with the only discrepancy due to the short
range of the potential cutoff.
   Annealing leads to an energy gain of approximately 0.13 eV per atom,
to give an energy in the final relaxed structure
0.28 eV higher than that of {\it c}-Si.
This stored energy is
approximately 6.1\% of {\it c}-Si cohesive energy and is primarily due
to bond angle distortions of around $16^\circ$ as shown in Fig.\ \ref{aangle}.

The 2194 K anneal causes partial melting of the amorphous material and
subsequent reconstruction into a higher density structure. The final quenched
structure contains more defects than the 1121 K annealed sample, but this
may be in part due to the rapid quench from a semi-liquid state. We observe
no indication of any recrystallization in the time simulated.
The 3000 K anneal clearly involves complete melting of the sample,
as the density increases to above that of crystalline Si. Again, the annealed
sample contains more defects than the 1121 K annealed sample, but less than
the un-annealed sample.

\section{Conclusions}
We have investigated the structural response of crystalline silicon to
ion-irradiation.
We are able to simulate stress generation and structural relaxation
with a relatively simple model.
Simulation of continual radiation followed by annealing
generates amorphous silicon with a
low level of defects.
   Amorphous silicon prepared by various experimental methods usually has a
density between 2\% and 10\% less than that of crystalline material,
whereas atomistic
simulations usually produce samples with a density around 4\% higher
than {\it c}-Si.
The fact that {\it a}-Si prepared by simulations of quenching liquid Si 
has a higher density than {\it c}-Si is not surprising. Amorphous silicon
has a structure similar to that of the liquid, which is 5\% denser than the
crystal form at the melting point \cite{ter88}.
Therefore a defect free continuous random network
(CRN)\cite{roo91} structure for {\it a}-Si would
be expected to result in a density greater than {\it c}-Si.

We conclude that
by conducting a direct simulation of radiation induced amorphization, we
have produced the experimentally observed
structure of ion beam amorphized Si.
The structure differs from the proposed defect free CRN models,
and differs from {\it a}-Si structures formed by simulations of quenching.
   Although this metastable
structure is at a higher energy than CRN {\it a}-Si, it cannot be annealed to
produce that structure, as transformation to {\it c}-Si, or melting
will occur at a lower temperature.
This leads us to question what is meant by `amorphous' Si, as an experimental
sample termed amorphous may contain micro-crystals, CRN structures, and
defects such as vacancies, interstitials, {\it c-a} boundaries, etc.
In some sense the never experimentally observed high density
structure can be regarded as true amorphous material,
whereas other less dense materials must be regarded
as containing defects at the very least, with a detailed structure that is
preparation dependent.

\section{Acknowledgments}
We thank M. Nastasi for illuminating discussions during
initial parts of this work.
This work was performed under the auspices of the United States Department
of Energy.

\appendix
\section{Derivation of the Virial for Many-Body Potentials}
\label{app1}

\subsection{Introduction}

   While the pressure virial is easy to calculate for a thermodynamic system
described by pair potentials, the extension to systems modeled by
many body potentials is not obvious.
Here we derive the virial for Tersoff type\cite{ter88} potentials,
but note that the method is readily
modified to other many body, or three body potentials so long as the
configuration energy can be written down as a function of atomic coordinates.

   Following Smith\cite{smi87} we write the pressure
with an explicit dependence on volume. This is achieved by relating absolute
atomic coordinates, ${\bf r}$, to scaled coordinates,
${\mbox{\boldmath $\rho$}}$ through the volume, $V$, so that
under isotropic expansion, or contraction of the system,
the scaled coordinates of atoms are unchanged:
\begin{equation}
{\bf r} = V^{1/3}{\mbox{\boldmath $\rho$}}.
\label{scale}
\end{equation}
   The pressure, $P$, is then given by:
\begin{equation}
P = NkT/V - \Phi/3V,
\label{press}
\end{equation}
where the virial, $\Phi$, is:
\begin{equation}
\Phi = 3V\langle \partial U([V^{1/3}{\mbox{\boldmath $\rho$}}]^N)/
\partial V\rangle .
\label{vir}
\end{equation}

\subsection{Application to the Tersoff Potential}

The configuration energy, $U$, is given by:
\begin{equation}
U = \sum_{i=1}^N\sum_{j<i} u_{ij}({\bf r}^N),
\label{ters}
\end{equation}
where $u_{ij}$ is the energy of the bond between atoms $i$ and $j$, as a
function of ${\bf r}^N$, the set of atomic coordinates.
The functional form is:
\begin{equation}
u_{ij}=f(r_{ij})[V_R(r_{ij})-b_{ij}V_A(r_{ij})],
\label{ters2}
\end{equation}
where $f(r_{ij})$ is
 a cutoff function that makes the interactions short
ranged, $V_R(r_{ij})$ and $V_A(r_{ij})$ are pair terms, $r_{ij}$ is the bond
length, and $b_{ij}$ is a many body term. As $b_{ij}$ is a function of the
positions of neighboring atoms of $i$ and $j$, these atoms experience a force
due to the $i$-$j$ bond.
Let the set of atoms involved in the calculation of $u_{ij}$,
i.e. atoms $i$, $j$ and neighbors, be denoted by $S_{ij}$.
Substituting Eq.\ (\ref{ters}) into Eq.\ (\ref{vir}) gives:
\begin{equation}
\Phi = 3V\langle\sum_{i=1}^N\sum_{j<i}\partial u_{ij}([V^{1/3}{\mbox{\boldmath $\rho$}}]^N)/
\partial V\rangle .
\label{virt}
\end{equation}
Since from Eq.\ (\ref{scale}):
\begin{equation}
\partial {\bf r} = {1\over3}V^{-2/3}{\mbox{\boldmath $\rho$}}\partial V ,
\label{dscale}
\end{equation}
this can be rewritten as:
\begin{equation}
\Phi = 3V\langle\sum_{i=1}^N\sum_{j<i}
{\partial u_{ij}({\bf r}^N)\over\partial{\bf r}^N}\cdot
{V^{1/3}\mbox{\boldmath $\rho$}^N\over 3V}\rangle .
\label{virt1a}
\end{equation}
Explicitly writing the dependence on interacting atoms gives:
\begin{equation}
\Phi = 3V\langle\sum_{i=1}^N\sum_{j<i}\sum_{k\in S_{ij}}
{\partial u_{ij}({\bf r}^N)\over\partial{\bf r}_k}\cdot
{V^{1/3}\mbox{\boldmath $\rho$}_k\over 3V}\rangle ,
\label{virt2}
\end{equation}
or, in terms of atomic positions:
\begin{equation}
\Phi = \langle\sum_{i=1}^N\sum_{j<i}\sum_{k\in S_{ij}}
-{\bf f}^{ij}_k\cdot
{\bf r}_k\rangle ,
\label{virt3}
\end{equation}
where ${\bf f}^{ij}_k$ is the force on atom $k$ due to the bond between
atoms $i$ and $j$.
 The atomic positions, ${\bf r}_k$, are local to the
bond being considered, e.g., periodic boundary conditions are accounted
for before the potential and force calculations.
 All local atomic coordinates
can in fact be translated to be relative to the position of atom $i$
for each bond,
so the force on atom $i$ can be neglected:
\begin{equation}
\Phi = \langle\sum_{i=1}^N\sum_{j<i}\sum_{k\in S_{ij}^\prime}
-{\bf f}^{ij}_k\cdot
{\bf r}_{ik}\rangle .
\label{virt4}
\end{equation}
  If the interaction
potential is pairwise, i.e., $b_{ij}$ depends only on atoms $i$ and $j$,
Eq.\ (\ref{virt4}) reduces to the pair virial:
\begin{equation}
\Phi = \langle\sum_{i=1}^N\sum_{j<i}
-{\bf f}^{ij}_j\cdot
{\bf r}_{ij}\rangle .
\label{virp}
\end{equation}

Individual elements of the pressure tensor, {\mbox{\boldmath $\Phi$}},
can be obtained by replacing
Eq.\ (\ref{scale}) with:
\begin{equation}
{\bf r} = {\bf L}{\mbox{\boldmath $\rho$}} ,
\label{scale2}
\end{equation}
where ${\bf L}$ is a $3\times 3$ matrix with determinant $V$.
For example, by using:
\begin{equation}
{\bf L} =
\pmatrix{ L_{xx} & 0 & 0 \cr
               0 & 1 & 0 \cr
               0 & 0 & 1 \cr} ,
\label{mat}
\end{equation}
one obtains:
\begin{equation}
\Phi_{xx} = 3\langle\sum_{i=1}^N\sum_{j<i}\sum_{k\in S_{ij}^\prime}
-f^{ij}_{x_k}
r_{x_{ik}}\rangle .
\label{virtxx}
\end{equation}

\protect\pagebreak

%
%

\begin{figure}
\vbox{
\includegraphics{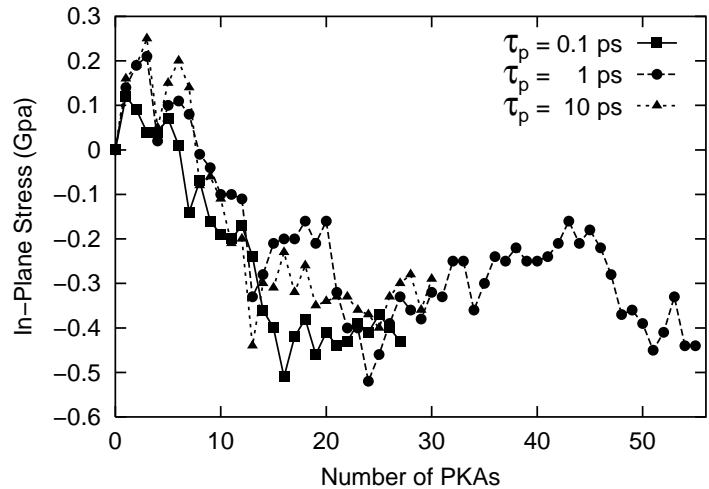}
\vspace{2.4 in}}

\caption{Variation in in-plane stress during amorphization.}
\label{stress2}

\end{figure}

%
%

\begin{table}
\caption{Structural properties of simulated samples.}
\label{props}
\begin{tabular}{lllldd}
Sample&
\multicolumn{1}{c}{Bond Length (\AA)}&
\multicolumn{1}{c}{Bond Angle ($^\circ$)}&
\multicolumn{1}{c}{Coordination}&
\multicolumn{1}{c}{Expansion (\%)}&
\multicolumn{1}{c}{Energy (eV/atom)}\\ \tableline
{\it c}-Si       &2.36$\pm$0.05&109.42$\pm$2.77&4.0&0.0&0.0\\
Amorphized:\\
$\tau_P$ = 0.1 ps&2.44$\pm$0.13&107.69$\pm$19.20&4.22$\pm$0.55&4.47&0.41\\
$\tau_P$ = 1.0 ps&2.45$\pm$0.14&107.58$\pm$19.44&4.27$\pm$0.54&1.91&0.41\\
$\tau_P$ =10.0 ps&2.45$\pm$0.14&107.44$\pm$19.76&4.32$\pm$0.54&$-$0.28&0.38\\
Annealed:\\
\ 300 K anneal    &2.44$\pm$0.13&107.60$\pm$19.40&4.26$\pm$0.54&2.12&0.37\\
1121 K anneal    &2.42$\pm$0.11&108.32$\pm$16.16&4.15$\pm$0.40&1.91&0.28\\
2194 K anneal    &2.43$\pm$0.12&107.95$\pm$17.60&4.23$\pm$0.45&$-$0.36&0.30\\
3000 K anneal    &2.43$\pm$0.13&107.77$\pm$18.26&4.26$\pm$0.48&$-$1.04&0.31\\
\end{tabular}
\end{table}

\end{document}